# Spectroscopy of divalent rare earth ions in fluoride crystals

R. Shendrik[1),2),*], A. S. Myasnikova[1)], E. A. Radzhabov[1),2)]. A. I. Nepomnyashchikh[1),2)]

1) Vinogradov Institute of geochemistry SB RAS (IGC SB RAS), Favorskii street 1a, Irkutsk, Russia, 664033

2) Physics faculty of Irkutsk state university, blvd Gagarina 20, Irkutsk, Russia, 664003

*Corresponding author: roshen@yandex.ru

**Abstract**

We have studied the absorption spectra of x-ray irradiation-induced $Ce^{2+}$ and $Pr^{2+}$ ions in crystals of alkaline-earth fluorides. We have calculated absorption spectra of divalent praseodymium ions in $SrF_2$ crystals doped with $Pr^{2+}$ for the first time. The calculated spectra agree rather well with the experimental data. In crystals containing induced $Ce^{2+}$ ions we have found strong electron-phonon coupling. In $BaF_2$, we do not observe bands corresponded to divalent Ce or Pr ions.

1. **Introduction**

Rare earths are found, usually, in the 3+ oxidation state. The energy difference between divalent and trivalent states is large and positive at the beginning lanthanide series indicating that the trivalent state is well-favored [1]. The negative or close to zero energy difference indicates that divalent state is more stable. It was observed for Eu, Yb and Sm ions. However, in crystals a certain fraction of any of the trivalent rare-earth ions can be reduced to the 2+ oxidation state by ionizing radiation [2-4], annealing the crystals in a reducing atmosphere of metal vapor, i.e. additive coloration [5, 6] or electrolytic means [7]. In reduced crystals of $CaF_2$-Ce and $CaF_2$-Pr divalent rare earth ions were observed [3, 4]. Furthermore, they were stable at room temperature. The reduced ions occupy alkaline-earth lattice sites having the full $O_h$ symmetry [3]. After reduction, the crystals are colored, typically having several broad, intense absorption bands extending from the near infrared through the UV region. Divalent cerium ions were also observed in $SrCl_2$-Ce crystals. The bands attributed to $Ce^{2+}$ were located in 400-700 nm region [8]. It has been found, that $Pr^{2+}$ ions take place in energy transfer in $SrF_2$-Pr and $CaF_2$-Pr scintillation crystals [9, 10]. Although wide investigations of $SrF_2$ and $BaF_2$ doped with Ce and Pr ions were provided, no divalent Ce and Pr ions were found in $SrF_2$ and $BaF_2$ crystals [3, 4, 5, 11].

The aim of our research is to derive from optical absorption spectra measurements and theoretical calculation an information about spin-orbital splitting and spectroscopy properties of divalent Ce and Pr ions in fluoride crystals. Part of the results was presented earlier at the paper [12]. Now a more elaborate overview and discussion of the results is given and recently obtained data are included.

## 2. Experimental methodology

Crystals of $CaF_2$ and $SrF_2$ were grown from the melt by the Bridgman-Stockbarger method in graphite crucibles in vacuum and were doped with 0.1 mol. % of $CeF_3$ or $PrF_3$. In alkaline earth fluoride single crystal growth a small amount of $CdF_2$ was generally used as a scavenger in order to remove oxides contained in the raw materials [12-15]. The crystals of $CaF_2$-Ce and $CaF_2$-Pr were irradiated at 300 K by x-rays from a Pd tube operating at 35 kV and 20 mA during not more than one hour. The crystals of $SrF_2$ were irradiated at 77 K.

The optical absorption spectra were obtained on a Perkin-Elmer Lambda 950 UV/VIS/NIR spectrophotometer at 7, 78, and 300 K at the Baikal Analytical Center for Collective Use, Siberian Branch of Russian Academy of Sciences. Absorption spectra at 7 K measured using cryocooler Janis Research CCS-100/204N.

The DFT (density functional theory) calculations were performed using embedded QM (quantum mechanical) cluster method implemented in GUESS computer code [17]. This method allows an integration of accurate quantum chemical calculations within a small defect-containing cluster and its nearest neighborhood with a classical description of the rest of crystal at molecular mechanics level. The QM cluster was surrounded by a large number (about 700 in this work) of atoms described classically with the shell model and pair potentials. About 70 cations between classical region and QM cluster were replaced with the interface atoms, which atoms have specially selected ECPs and pair potentials to minimize the distortion of electronic and spatial structure at the QM cluster edge. All atoms of the classical, interface and QM regions are allowed to relax during the geometry optimization, thus simulating the deformation and polarization of the crystal lattice by the presence of the impurity ions. For the classical region we have used the pair potentials parameters in the Buckingham form for $CaF_2$ and $SrF_2$ crystals from Ref. [18] and adjusted them so that they are compatible with GUESS code. The classical region was surrounded by several thousand of fixed-point charges to simulate the correct Madelung potential of the crystal. The GUESS code is not capable to calculate the electronic structure of QM cluster, but calls another quantum-chemical program (in this case, the Gaussian 03 [19]), and adds classic terms to find the total energy of the system. The applicability of embedded cluster calculation method for defects in ionic crystals is described in Refs. [20, 21] in more details.

Calculations have been performed in a cluster containing 12 calcium or strontium ions (Me) and 32 fluorine ions, the praseodymium ion was placed in the central position instead of alkali ion - $[Me_{12}F_{32}Pr^{2+}]$-6 (Fig. 5). The 6–31G* basis sets on fluorine and calcium ions, and SDD basis set on strontium and praseodymium ions were used. All cations in 6 Å vicinity of QM cluster were replaced by the interface ions, which had LANL1 ECP (Hay-Wadt large core pseudopotential) and no basis functions.

For the DFT calculations we used the modified B3LYP functional containing 40% of Hartree-Fock and 60% of DFT exchange energies which showed most adequate electron state localization and was successfully employed for DFT calculations of defects in fluoride crystals [20, 21]. Optical energies and dipole matrix elements of transitions were calculated with the time-dependent DFT (TDDFT) method applied.

Note that in the previous papers [12, 21] we carried out the calculations of the spatial structure and optical properties of $Pr^{3+}$ ion in crystals of alkaline earth fluorides. The calculations have been performed on the above scheme, and showed good agreement with experiment data. This paper presents the results of calculations of $Pr^{2+}$ single center using $[Ca_{12}F_{32}Pr^{2+}]^{-6}$ and $[Sr_{12}F_{32}Pr^{2+}]^{-6}$ cluster.

### 3. Experimental results

*A. $CaF_2$-$Ce^{2+}$ and $SrF_2$-$Ce^{2+}$*

After x-ray irradiation, absorption bands in a wide spectral region appear in $CaF_2$-Ce and $SrF_2$-Ce crystals. They are due to appear of photochromic centers [16] and divalent Ce, when trivalent Ce is reduced by x-ray irradiation. In $CaF_2$ these centers remain relatively stable at room temperature. In $SrF_2$ crystals, divalent Ce is stable only at low temperatures. Most of the $Ce^{3+}$ ions remains unconverted after irradiation.

Figure 1 shows the absorption spectra divalent Ce in $CaF_2$ and $SrF_2$ crystals. At low temperatures, the absorption begins at around 7070 cm$^{-1}$ in $CaF_2$-Ce and at around 6400 cm$^{-1}$ in $SrF_2$-Ce and fills the whole region of the spectra. Groups of relatively sharp lines are observed at about 7070 cm$^{-1}$ and 8800 cm$^{-1}$ in $CaF_2$-Ce, and at about 6400 cm$^{-1}$ and 8080 cm$^{-1}$ in $SrF_2$-Ce. The higher energy bands except two sharp lines at 26400 cm$^{-1}$ and 26800 cm$^{-1}$ (inset to the fig. 1a) are broadened.

The spectrum exhibits a very strong temperature dependence that is shown in Figure 2 and Figure 3. In $CaF_2$-Ce, at 300 K and in $SrF_2$-Ce at 77 K the various components can hardly be resolved. However, at low temperatures some of lines are very sharp. Temperature dependences of line width of bands at 7070 cm$^{-1}$ and 8800 cm$^{-1}$ in CaF2 and 6400 cm$^{-1}$ and 8080 cm$^{-1}$ in $SrF_2$ are given in Figure 3. The line width of the narrowest band at 7 K is limited of spectrophotometer resolution.

*B. $CaF_2$-$Pr^{2+}$ and $SrF_2$-$Pr^{2+}$*

After x-ray irradiation, absorption bands in 9000 – 35000 cm$^{-1}$ in CaF$_2$-Pr, and in 8000-24000 cm$^{-1}$ in SrF$_2$-Pr appear. They are related to divalent Pr. In CaF$_2$, divalent Pr centers are relatively stable at room temperature, similar to Ce$^{2+}$. However, in SrF$_2$ crystals divalent Pr is stable only at low temperatures. Temperature dependence of stability divalent Pr ions is given in the inset to Figure 4.

Figure 4 shows absorption spectra of divalent Pr in CaF$_2$ and SrF$_2$. In CaF$_2$ three groups of broad bands at about 8900, 16000, and 25000 cm$^{-1}$ can be distinguished (fig. 4, a). In SrF$_2$ two groups of broad bands at about 9000 and 20000 cm$^{-1}$ are found (fig. 4, b). In BaF$_2$ we do not observe bands corresponded to divalent Ce or Pr ions.

### 4. Discussion

The sharp lines in the absorption spectrum of CaF$_2$-Ce$^{2+}$ are related to 5d-4f transitions in Ce$^{2+}$ [3]. In fig. 1a vertical lines in top of the figure are energy levels of spin-orbital splitted 4f$^2$ state of free Ce$^{2+}$ ion. We observed a good agreement between these levels and sharp line positions. Observed lines in absorption are related to transition from ground 4f$^1$5d$^1$ state to 4f$^2$ level terms. Sharp lines at 26400 cm$^{-1}$ and 26800 cm$^{-1}$ can be attributed to 4f$^1$5d$^1$->4f$^0$5d$^2$ transitions. In SrF$_2$-Ce the spectrum has the same structure. Thus, the groups of sharp lines at 6420 cm$^{-1}$ and 8080 cm$^{-1}$ are related to 4f$^1$5d$^1$->4f$^2$($^3$H$_4$) and 4f$^1$5d$^1$->4f$^2$($^3$H$_5$) transition in Ce$^{2+}$ ion. At low temperature most of the intensity of the transition is in single line. The strong temperature dependence is certainly associated with populating only the lowest vibrational level of 5d state from which a single transition is observed to very closely lying vibrational levels of the 4f configuration.

From the other hand, the electron-phonon coupling strength for Ce$^{2+}$ ions can be probed by temperature dependent line broadening. At elevated temperatures, the line width of the zero-phonon lines increases due to phonon induced relaxation processes. The line width of a zero-phonon line is determined by the lifetime of the starting and final level. The total line width depends on the temperature ($T$) and is a summation of several contributions, for example one-phonon emission and absorption, two-phonon Raman processes, multiphonon relaxation and Orbach processes [22, 23]. Temperature dependent line width measurements can provide information on the electron-phonon coupling strength since stronger electron-phonon coupling will result in more broadening.

Significant contributions to the line width give two processes. The first one is the contribution due to the direct or one-phonon process, where one phonon is emitted or absorbed. Temperature dependence of line broadening in the case of this contribution is described by [22]:

$$\Gamma^d(T) \sim \coth\left(\frac{\hbar\omega}{2k_BT}\right), \quad (1)$$

where $\Gamma^d(T)$ is the spectral width of the band at temperatures $T$, $\omega$ is phonon frequency, $k_B$ is Boltzmann constant.

Two-phonon Raman relaxation processes also exists. In the Raman two-phonon relaxation, a certain level $i$ absorbs a phonon that bridges the gap to an intermediate level $j'$; the system relaxes with a higher energetic phonon to level $j$. The Raman two-phonon process is a nonresonant process. Using the Debye approximation of phonon energies, it can be derived that the temperature-dependent contribution to the line width by the Raman process is described by [22 – 24]

$$\Gamma^R(T) = \alpha \left(\frac{T}{T_D}\right)^7 \int_0^{\frac{T_D}{T}} \frac{x^6 e^x}{(e^x-1)^2} dx, \quad (2)$$

In this equation $\alpha$ is the electron-phonon coupling parameter for the Raman process, $T_D$ is the effective Debye temperature and $x = \frac{\hbar\omega}{k_BT}$. The contribution of the direct process is relatively large at lower temperatures, but at higher temperatures the Raman process becomes dominant [22–24]. In Fig. 3, the temperature dependences of the line width from $4f^15d^1$ to $^3H_4$ and $^3H_5$ transitions are shown. The drawn lines are fits to the equation:

$$\Gamma(T) = \Gamma^0 + \Gamma^R(T). \quad (3)$$

Here $\Gamma^0$ is the line width at 7 K. Direct phonon broadening (Eq. 1) and Fano broadening due to close to conduction band position of $4f^15d^1$ levels in divalent Ce ion cause it. The $\alpha$ values obtained for the studied transitions in $CaF_2$ are for $4f^15d^1 \rightarrow {}^3H_4$ is 686 cm$^{-1}$ and for $4f^15d^1 \rightarrow {}^3H_5$ is 914 cm$^{-1}$. In $SrF_2$-$Ce^{2+}$ we observed the same behavior of $Ce^{2+}$ line broadening. The $\alpha$ values obtained for the transitions in $SrF_2$ are for $4f^15d^1 \rightarrow {}^3H_4$ is 1138 cm$^{-1}$ and for $4f^15d^1 \rightarrow {}^3H_5$ is 1438 cm$^{-1}$. Electron-phonon coupling strengths for $Ce^{2+}$ are much higher than for $Ce^{3+}$ (typical values about 400 cm$^{-1}$) [24]. It is the evidence of strong electron-phonon interaction with $Ce^{2+}$ ions. In $SrF_2$, this interaction is stronger than in $CaF_2$.

Position of $^3H_4$ and $^3H_5$ terms related to the bottom of conduction band can be estimated according to photostimulated luminescence data given in [25]. $^3H_4$ term is below the bottom of conduction band to 15000 cm$^{-1}$ in $CaF_2$ and 13000 cm$^{-1}$ in $SrF_2$. The narrowest lines corresponding to transition $4f^15d^1 \rightarrow {}^3H_4$ in $CaF_2$ and $SrF_2$ have $\Gamma$ about 0.9 cm$^{-1}$ and 1.8 cm$^{-1}$ correspondingly. $^3H_5$ term is below the bottom of conduction band to 8000 and 6500 cm$^{-1}$ in $CaF_2$ and $SrF_2$ correspondingly. At once, the lines related to $4f^15d^1 \rightarrow {}^3H_5$ transition are more broadened. Those $\Gamma$ are about 6.1 cm$^{-1}$ in $CaF_2$ and 29 cm$^{-1}$ in $SrF_2$. $^3H_5$ levels of $Ce^{2+}$ in $CaF_2$ and $SrF_2$ embedded in the conduction band continuum, thus one expect this level to autoionize. Such a process should lead also to a Fano type broadening of the line. This process takes place although the inner 4f electrons, which are shielded, by the $5s^1$ and $5p^6$ outer electrons, and this shielding ionize also much less readily [26].

We observe the strong electron-phonon interaction and the sidebands which accompanies the lines demonstrate vibronic origin. The frequencies of a longitudinal optic phonon obtained from optical data at 5 K were 484 cm$^{-1}$ and 397 cm$^{-1}$ for $CaF_2$ and $SrF_2$, respectively [27]. Typical distances between the lines in the absorption spectra are 200 and 160 cm$^{-1}$ for $CaF_2$-$Ce^{2+}$ and $SrF_2$-$Ce^{2+}$, respectively. Estimated frequencies of these local vibrations are lower than the lattice phonon frequencies in $CaF_2$ and $SrF_2$ crystals. It appears due to strong distortion of the lattice near $Ce^{2+}$ ion, because $Ce^{2+}$ is substantially larger than $Ce^{3+}$ [28]. There is good agreement between these vibronic frequencies observed with Raman spectroscopy in $CaF_2$-$La^{2+}$ [29] and those obtained from the vibronic side bands. From the other hand, the observed lines can be corresponded to crystal field splitting of the final 4f level. The crystal field splitting is of the same order of magnitude as the vibrational energies. Overlap between vibronic lines and the strong zero phonon lines due to transitions to different crystal field components makes it hard to identify vibronic transitions.

*B. Pr-doped crystals*

In absorption spectrum of $CaF_2$-$Pr^{2+}$ we observe three bands at 8900, 16000, and 25000 cm$^{-1}$. In contrast to $CaF_2$-$Ce^{2+}$, these bands do not exhibit sharp structure at 7 K. In previous paper, we have shown that these bands are related to 4f-5d transitions in $Pr^{2+}$ ion. The 25000 cm$^{-1}$ band is attributed to the threefold degenerate t-level and the 8900 cm$^{-1}$ band to the doublet e-level. The absorption spectrum of $SrF_2$-$Pr^{2+}$ has the same structure. We observed wide bands at 9000 and 20000 cm$^{-1}$.

We performed theoretical calculation of optical absorption spectra to identify these bands. As a first step, we calculated the lattice distortion induced by presence of $Pr^{2+}$ ion in $CaF_2$ and $SrF_2$ crystalline matrices. The praseodymium ion induces considerable lattice distortions, so the eight fluorine ions nearest to the $Pr^{2+}$ ion displace away from it by 0.082Å for $CaF_2$ and 0.023Å for $SrF_2$ crystalline matrices, which are 3.44% and 0.91% from the $Pr^{2+}$- F distance in unrelaxed cluster respectively. All twelve cations displace toward $Pr^{2+}$ ion by 0.046Å (1.19%) and 0.028Å (0.68%) for $CaF_2$ and $SrF_2$ respectively. The geometry optimization shows that the displacements of the ions are much larger for the $CaF_2$ crystal. This fact is due to a large ionic radii of $Sr^{2+}$ as compared with $Pr^{2+}$ ions (1.27 and 1.19 Å, respectively) and smaller ionic radius of $Ca^{2+}$ (1.04 Å).

Using the geometry of the ground state, we performed the calculation of $4f^3 \rightarrow 4f^2 5d$ absorption spectra. The optical absorption spectra calculated with TDDFT are shown in Fig. 5 (d). The smooth curve shown in the figure is a convolution of Gaussian-type functions, each centered at the excitation energy of the corresponding transition and weighed with the value of the corresponding oscillator strength. The full width at half maximum (FWHM) was selected empirically for the best display of calculated results. In the central configuration of $Pr^{2+}$ ion the cubic crystal field splits 5d states into triply degenerate components $t_{2g}$ and doubly degenerated components $e_g$. So for the single ion calculation only two main bands were observed in absorption spectra.

The lowest and highest absorption bands are connected with the transitions from 4f state to $e_g$ and $t_{2g}$ levels of 5d states of $Pr^{2+}$ ion respectively. The shape and position of this band agree well with the experimental absorption bands. Note that the low energy shift of calculated absorption bands is due to different bases on calcium and strontium ions. The third absorption band is centered at 12900 cm$^{-1}$ (1.6 eV), which is also in agreement with the experimentally observed band at 16000 cm$^{-1}$ (1.9 eV). The one-state electron states participating in the transition consist of s-states of different ions (Fig. 5) One-electron interpretation of this band is difficult due to the electron state participating in the transition comprises s-states of surrounding fluorine ions. Furthermore, in magnet circular dichroism behavior of this band is different to behavior other $Pr^{2+}$ bands [30].

Ab initio calculations also show that 20000 cm$^{-1}$ band is attributed to the transition from ground 4f state to the threefold degenerate t-level, and 9000 cm$^{-1}$ to doublet e-level. In table 1 energies of the 4f-5d transitions and energy of crystal-field splitting of 5d state are given. Well agreement between experimental and theoretical data are observed.

The spectra were calculated without the spin-orbit interactions. Spin-orbital splitting of 5d state is much lower than crystal field splitting. Therefore, the spin-orbit interaction is not much influence on the calculation. In the previous paper [21] we carried out the calculations of the spatial structure and optical properties of $Ce^{3+}$ ion in crystals of alkaline earth fluorides. The calculations have been performed on the above approach, and showed good agreement with experimental data and theoretical calculations with spin-orbital interaction [31].

Bands corresponded to 4f-5d transitions in the $Pr^{2+}$ doped crystals are broadened. This fact can be explained as follows. 5d($e_g$) level of $Pr^{2+}$ ion is located close to conduction band of the host crystal [32]. In this case, autoionization process can also lead to broadening of all the optical transitions and smear out any fine structure in the absorption [33, 34].

$Pr^{2+}$ ion in $SrF_2$ crystal is unstable at room temperature. Its thermal quenching is started at 130 K (Fig. 4, inset) and $Pr^{2+}$ ions are almost absent at 160 K. Temperature dependence of optical density of absorption bands corresponded to $Pr^{2+}$ ions is agreed well with estimation of temperature stability of $Pr^{2+}$ ions based on x-ray luminescence and thermoluminescence data. This fact confirms the proposed energy transfer mechanism in $SrF_2$-Pr crystals [9].

**6. Conclusions**

We found $Ce^{2+}$ and $Pr^{2+}$ ions in irradiated $CaF_2$ and $SrF_2$ crystals doped with Ce or Pr ions. In $CaF_2$ the divalent rare earth are stable at room temperature, but in $SrF_2$ ions of $Pr^{2+}$ are stable only at lower temperatures. We observe three bands in the absorption spectrum corresponding to 4f-5d transitions in $CaF_2$-$Pr^{2+}$ and two bands in $SrF_2$-$Pr^{2+}$. Experimental data are agreed well with calculated spectra.

In Ce-doped crystals after irradiation, we observed bands corresponding to $4f^15d^1$-$4f^2$ transitions in a $Ce^{2+}$ ion that are well resolved at low. Similar to Pr-doped crystals, the $Ce^{2+}$ ions are stable at room temperature in $CaF_2$. Spectral lines corresponding to $4f^15d^1$->$4f^2(^3H_4)$ and $4f^15d^1$->$4f^2(^3H_5)$ transitions have vibronic structure. Calculated from temperature dependence of line width electron-phonon coupling strengths for $Ce^{2+}$ are much higher than for $Ce^{3+}$. It is the evidence of strong electron-phonon interaction with $Ce^{2+}$ ions in fluorides.

**Acknowledgment**

The reported study was funded by RFBR according to the research project No. 15-02-06666_a.

**List of figures**

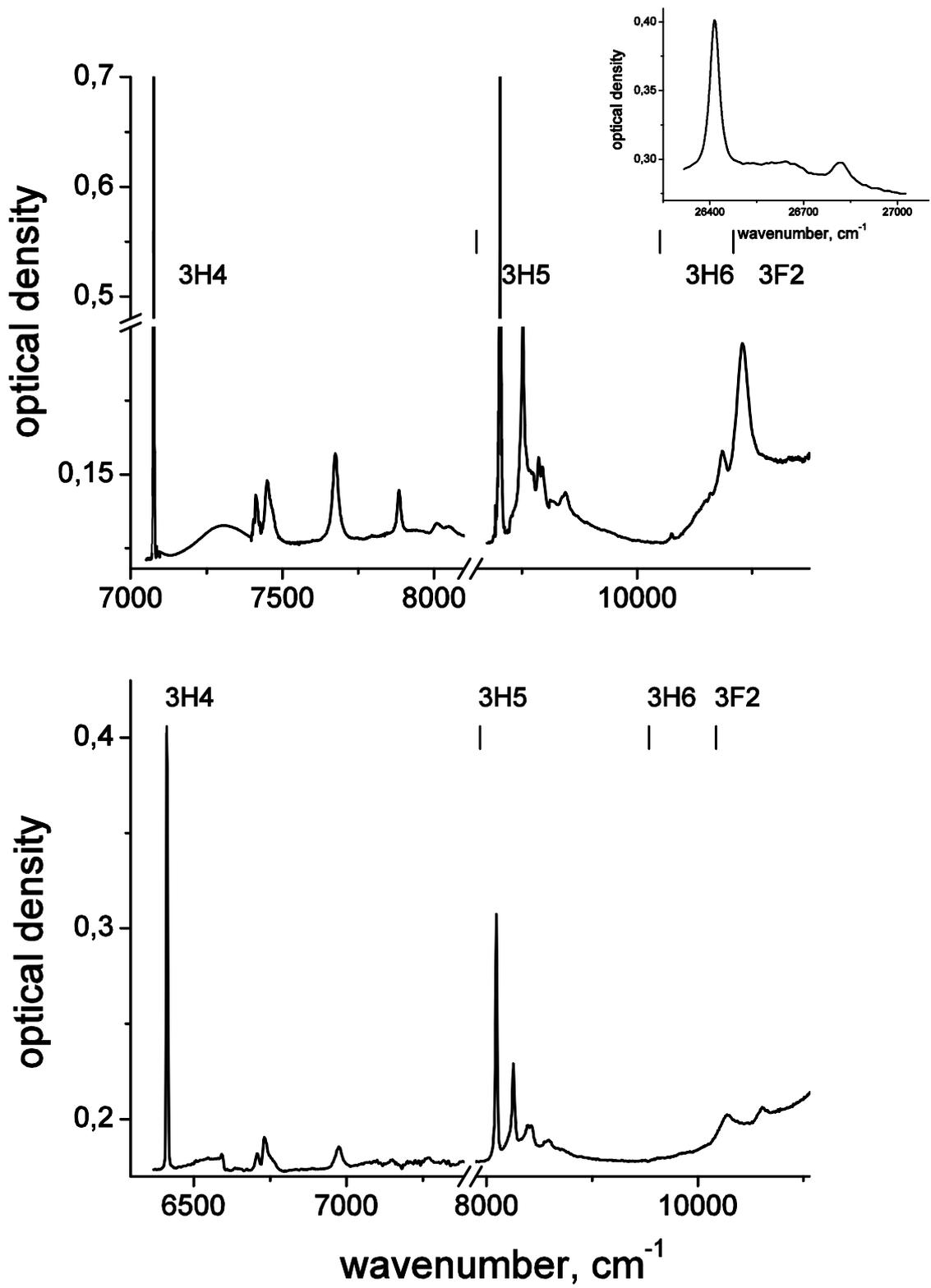

Figure 1. Absorption spectra of $CaF_2$-$Ce^{2+}$ (a) and $SrF_2$-$Ce^{2+}$ (b) at 7 K. Energies of spin-orbital splitted 4f level of free $Ce^{2+}$ given in the top of the figures. In the inset of figure (a) sharp lines in UV spectral region corresponding to $Ce^{2+}$ are given.

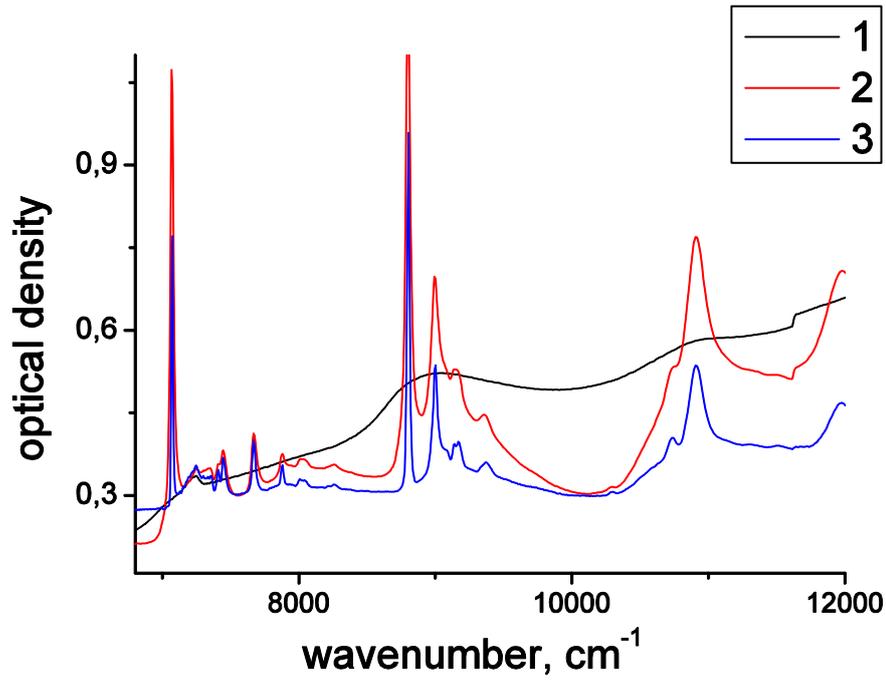

Figure 2. Absorption spectrum of $CaF_2$-$Ce^{2+}$ at 300 K (1), 78 K (2), and 7 K (3).

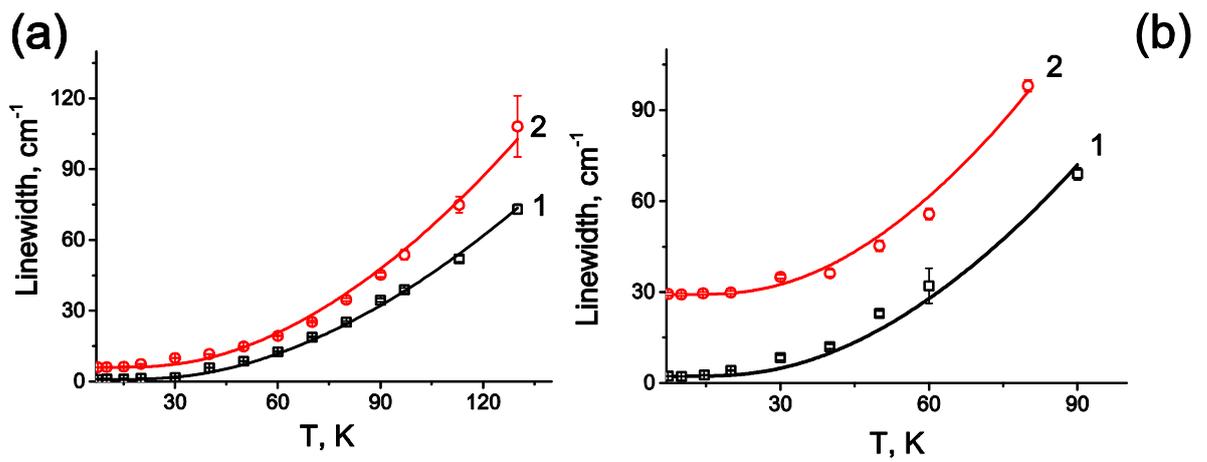

Figure 3. Temperature dependence of line width of the absorption bands at 7070 cm$^{-1}$ (curve 1, squares) and 8800 cm$^{-1}$ (curve 2, dots) in $CaF_2$-$Ce^{2+}$ (a) and the absorption bands at 8080 cm$^{-1}$ (curve 1, squares) and 6400 cm$^{-1}$ (curve 2, dots) in $SrF_2$-$Ce^{2+}$ (b). The lines show the result of fitting.

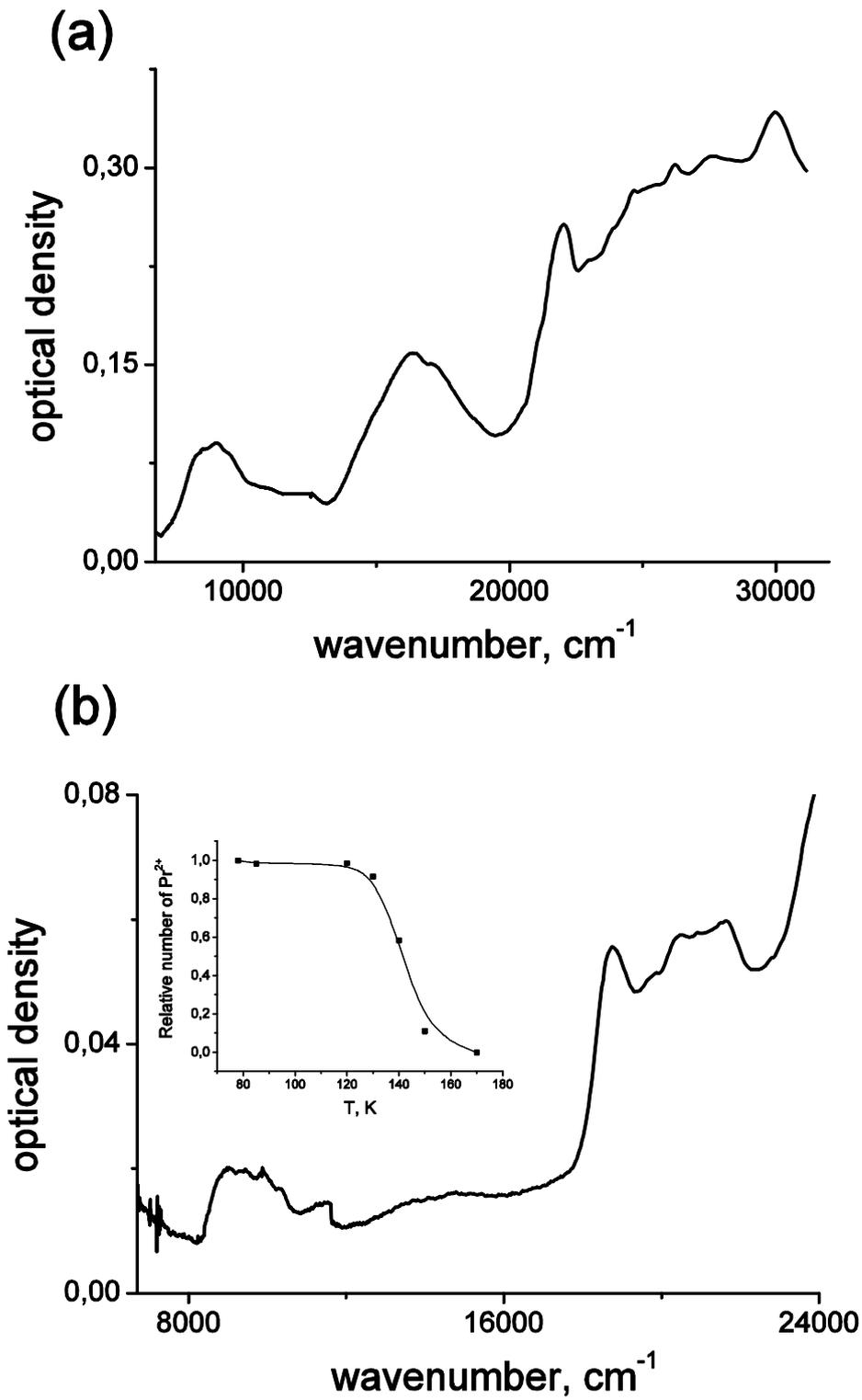

Figure 4. Absorption spectra of $CaF_2$-$Pr^{2+}$ at 7 K (a) and $SrF_2$-$Pr^{2+}$ at 78 K (b). In the inset (b), temperature dependence of optical density of bands attributed to $Pr^{2+}$ ions in $SrF_2$ is given.

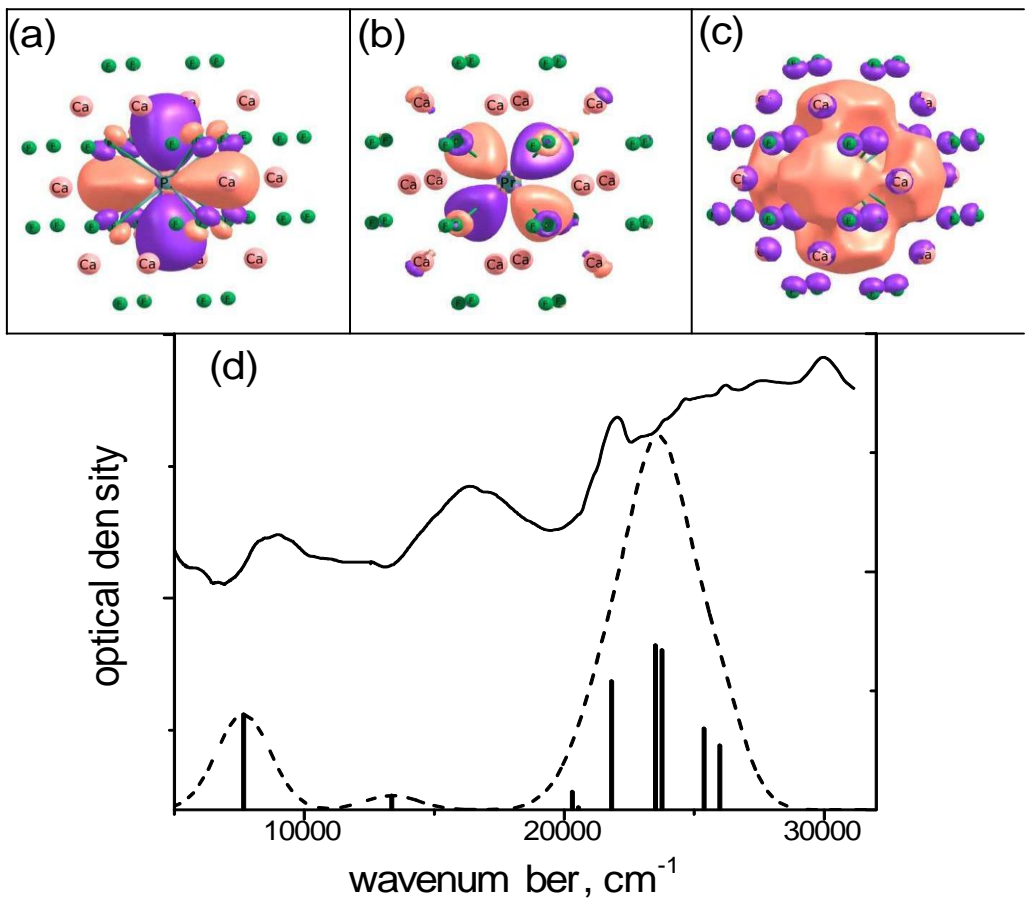

Figure 5. One-electron states participating in the most intense optical absorption transitions: a) – band at 7250 cm$^{-1}$ (0.9 eV), b) – band at 21778 cm$^{-1}$ (2.7 eV), c) – band at 12900 cm$^{-1}$ (1.6 eV) – and comparison between calculated and experimental absorption spectra in CaF$_2$-Pr$^{2+}$ crystal.

**Table 1.** Comparison calculated and experimental energies of absorption peaks corresponding to 4f-5d($e_g$) transitions and crystal-field splitting for $Pr^{2+}$ ions in $CaF_2$ and $SrF_2$ crystals.

|  | $CaF_2$-$Pr^{2+}$ | | $SrF_2$-$Pr^{2+}$ | |
|---|---|---|---|---|
|  | 4f->5d ($e_g$) [cm$^{-1}$] | 5d splitting[*)] [cm$^{-1}$] | 4f->5d ($e_g$) [cm$^{-1}$] | 5d splitting[*)] [cm$^{-1}$] |
| **Experimental** | 8800 | 10900 | 9195 | 9700 |
| **Calculation** | 8360 | 12500 | 8600 | 14000 |

[*)] Energy of 5d state crystal field splitting (i.e. energy gap between $e_g$ and $t_{2g}$ levels of 5d-state)